\documentclass[12pt]{article}
\usepackage{amsfonts,amssymb,amsmath}
\def\dac{\displaystyle\frac}
\def\d{\partial}
\def\[{\left[}
\def\]{\right]}
\def\({\left(}
\def\){\right)}

\newcommand{\diag}{\mathop{\rm diag}\nolimits}

\begin{document}

\begin{center}
I. V. Kirnos\\
\textit{Tomsk State University of Control Systems and Radioelectronics}\\
ikirnos@sibmail.com, kirnos@tspu.edu.ru\\
\Large\textbf{Some Cosmological Solutions in an Arbitrary Order of Lovelock Gravity}
\end{center}

\begin{abstract}
We consider the Lovelock theory of gravity that assumes a nonlinearity
of the field equations in the second-order derivatives of the metric.
We prove the opportunity of obtaining cosmological solutions without
isotropization in the presence of matter in the form of a perfect fluid,
which is necessary for invisibility of extra dimensions that inevitably
emerge in the Lovelock theory. In particular, the Jacobs solution has been
generalized to an arbitrary order of the theory, and in the third order,
an anisotropic exponential solution has been found.
\end{abstract}

The recent observations of supernovae in distant galaxies and CMB anisotropy indicate
the incorrectness of the standard cosmological model and require a \``new cosmology\''
which can possibly rest on a new theory of gravity. One of such possible theories is the
Lovelock theory which consists in the following.

In general relativity (GR) the gravitational field equations have the form $G_{\mu\nu}=\varkappa T_{\mu\nu}$,
where $T_{\mu\nu}$ is the energy-momentum tensor, $\varkappa$ is a constant factor, while the gravitational
field tensor $G_{\mu\nu}$ satisfies the following conditions:
\begin{enumerate}
\item Symmetry: $G_{\mu\nu}=G_{\nu\mu}$.
\item Zero divergence: $\nabla_\mu {G^\mu}_\nu=0$.
\item A dependence only on the metric, its first and second order derivatives:\\
$G_{\mu\nu}=G_{\mu\nu}(g_{\alpha\beta},\d_\gamma g_{\alpha\beta},\d_\gamma \d_\delta g_{\alpha\beta})$.
\item Linearity in the second-order derivatives of the metric tensor.
\end{enumerate}
The general form of such a tensor is
$$G_{\mu\nu}=\alpha(R_{\mu\nu}-\dac 1 2 Rg_{\mu\nu})+\Lambda g_{\mu\nu},$$
where $\alpha$ and $\Lambda$ are arbitrary constants. We thus obtain the well-known
Einstein-Hilbert equations with a cosmological constant.

Lovelock \cite{lovelock} suggested to abandon the condition of linearity
with respect to second-order derivatives and to preserve only the requirements
of symmetry, zero divergence, and the dependence on the metric and its derivatives.
In this case we obtain:
$${G^\mu}_\nu=\sum_{p=0}^{\left[\frac{n-1}{2}\right]}\alpha_p\delta^{\mu\lambda_1\lambda_2\cdots\lambda_{2p}}_{\nu\sigma_1\sigma_2\cdots\sigma_{2p}} {R^{\sigma_1\sigma_2}}_{\lambda_1\lambda_2} {R^{\sigma_3\sigma_4}}_{\lambda_3\lambda_4}\cdots {R^{\sigma_{2p-1}\sigma_{2p}}}_{\lambda_{2p-1}\lambda_{2p}},$$
where $n$ is the space-time dimensionality, square brackets denote the integer part of a fraction,
$\alpha_p$ are arbitrary constant factors, and $\delta^{\cdots}_{\cdots}$ is the multi-index delta symbol
(equal to $\pm 1$ if the upper indices form an even or odd permutation of the lower ones, respectively,
and zero in all other cases). It is sometimes helpful to rewrite this sum as
$${G^\mu}_\nu=\sum_{p=0}^{\left[\frac{n-1}{2}\right]}\alpha_p {G^{(p)\mu}}_\nu,$$
where evidently
$${G^{(p)\mu}}_\nu=\delta^{\mu\lambda_1\lambda_2\cdots\lambda_{2p}}_{\nu\sigma_1\sigma_2\cdots\sigma_{2p}} {R^{\sigma_1\sigma_2}}_{\lambda_1\lambda_2} {R^{\sigma_3\sigma_4}}_{\lambda_3\lambda_4}\cdots {R^{\sigma_{2p-1}\sigma_{2p}}}_{\lambda_{2p-1}\lambda_{2p}}.$$

It is necessary to point out a peculiarity of Lovelock's theory: in the case of usual 4-dimensional
space-time, $n=4$, we obtain a sum from zero to unity, that is, we return to GR. Thus to obtain
something new in Lovelock gravity, one has to consider spaces with dimensions not less than five.
But since we do not observe any extra dimensions, we have to deal with anisotropic spaces and seek anisotropic cosmological solutions in Lovelock's theory.

Before directly describing solutions in Lovelock's theory, let us look at the situation in GR.
If we consider a flat anisotropic model,
$$g_{\mu\nu}=\diag\{-1,a_1^2(t),a_2^2(t),\ldots,a_n^2(t)\},$$
(here, for convenience, we denote by $n$ the number of spatial dimensions, without the time coordinate), then, in the absence of matter, we obtain the Kasner solution which can be represented as
$$g_{\mu\nu}=\diag\{-1,t^{2p_1},t^{2p_2},\ldots,t^{2p_n}\},\quad \sum_i p_i=1,\quad \sum_{i<j}p_ip_j=0;$$
in the presence of matter with the maximally stiff equation of state, we have the Jacobs solution
$$g_{\mu\nu}=\diag\{-1,t^{2p_1},t^{2p_2},\ldots,t^{2p_n}\},\quad \sum_i p_i=1,\quad \sum_{i<j}p_ip_j=-\dac{\varkappa}{2}\varepsilon_0,$$
where $\varkappa=8\pi G/c^4$, $\varepsilon_0$ is the initial energy density; with any other
value of the equation-of-state parameter we obtain a solution which at late times approaches
an isotropic one. It can mean that the extra spatial dimensions become equivalent to the
usual ones and consequently observable.

In GR isotropization is a good feature because it explains the observed spatial isotropy.
But in Lovelock's theory there are extra dimensions that should not become equivalent to
the main ones, therefore isotropization of \textit{all} spatial dimensions is undesirable.
If this happens in this theory, then we should either abandon the theory itself or invoke
an anisotropic matter with different pressures in different directions, or invent some other
complications. Fortunately, it turns out that all this is unnecessary. To understand that,
let us see what happens in the second order of Lovelock's theory.

If we consider the second order together with the first one, it turns out that power-law
solutions are absent. Therefore we discard the first order and address only the second one:
$$\alpha G^{(2)}_{\mu\nu}=\varkappa T_{\mu\nu}.$$
Then, in the absence of matter, we obtain an analogue of Kasner's solution discovered
by Deruelle \cite{deruelle} and later re-discovered by Toporensky and Tretyakov \cite{toporensky-tretyakov}:
$$\sum_i p_i=3,\quad \sum_{i<j<k<l}p_ip_jp_kp_l=0.$$
In the presence of matter of a certain type (namely, $p=w\rho$, $w=1/3$), we obtain
an analogue of the Jacobs solution found some time ago by the author
\cite{jacobs-2-yalchik,kirnos-makarenko-pavluchenko-toporensky}:
$$\sum_i p_i=3,\quad \sum_{i<j<k<l}p_ip_jp_kp_l=-\dac{\varkappa\varepsilon_0}{96\alpha}.$$

Writing down all these solutions on a single page, it is easy to notice their large similarity and
even to suppose how the corresponding solutions can look in an arbitrary order.

It is indeed possible to find such solutions. In the $p$-th order of Lovelock's theory
$$\alpha_p G^{(p)}_{\mu\nu}=\varkappa T_{\mu\nu}$$
we obtain the generalized Kasner solution:
$$\sum_i p_i=2p-1,\quad \sum_{i_1<i_2<\cdots<i_{2p}}p_{i_1}p_{i_2}\cdots p_{i_{2p}}=0,$$
and the generalized Jacobs solution
$$\sum_i p_i=2p-1,\quad \sum_{i_1<i_2<\cdots<i_{2p}}p_{i_1}p_{i_2}\cdots p_{i_{2p}}=-\dac{\varkappa\varepsilon_0}{2^p(2p)!\alpha_p},$$
corresponding to the equation-of-state parameter $w=1/(2p-1).$ After obtaining these solutions
I came to know that the generalized Kasner solution and even the above value of the parameter $w$
were found earlier by Pavluchenko \cite{pavluchenko}. However, for some reason, he did not obtain
the generalized Jacobs solution, which is presented here.

One should now say a few words on the shortcoming of this solution. Yes, it proves the
opportunity of obtaining an expanding space without isotropization, but only for a single value of $w$.
One can suspect that in each order of Lovelock's theory there is only one distinguished value of $w$
at which a non-isotropizing model is possible, while at all other $w$ the model isotropizes.
Moreover, while obtaining this solution, we discarded all terms of not only higher orders than $p$
in the curvature (which can be simply caused by the dimensionality of space) but also all lower orders.
It can be done at early stages of the expansion when the curvature is large. But what happens when
it becomes small?

Another form of solution, the exponential one, makes it possible to answer this question;
it also allows one not to neglect the lower orders. Some time ago such a solution was obtained
for the second order \cite{jacobs-2-yalchik}: the equations
$$G^{(1)}_{\mu\nu}+\alpha_2 G^{(2)}_{\mu\nu}=\varkappa T_{\mu\nu}$$ have the solution
$$g_{\mu\nu}=\diag\{-1,e^{2H_1t},e^{2H_2t},\ldots,e^{2H_nt}\},$$
with $$\sum_i H_i=0,\quad\sum_i H_i^2=(1-3w)\varkappa\varepsilon_0,\quad\sum_i
H_i^4=\dac{w-1}{2\alpha_2}\varkappa\varepsilon_0+\dac{(1-3w)^2}{2}\varkappa^2\varepsilon_0^2,$$
and the parameter $w$ is here to a large extent arbitrary, which refutes the suspicion on the existence
of a distinguished value of $w$.

Now we have considered the third-order equations
$$\alpha_1 G^{(1)}_{\mu\nu}+\alpha_2 G^{(2)}_{\mu\nu}+\alpha_3  G^{(3)}_{\mu\nu}=\varkappa T_{\mu\nu}$$
and obtained an exponential solution which has a more complex form,
$$g_{\mu\nu}=\diag\{-1,e^{2H_1t},e^{2H_2t},\ldots,e^{2H_nt}\},$$
with $$\begin{array}{l}\sigma_1=0,\\ \sigma_4=\dac 1 2
\sigma_2^2-\dac{\alpha_1}{2\alpha_2}\sigma_2-\dac{1-5w}{16\alpha_2}\varkappa\varepsilon_0,\\
\sigma_6=\dac 1 3 \sigma_3^2+\dac 1 4
\sigma_2^3-\dac{3\alpha_1}{8\alpha_2}\sigma_2^2+\dac{1}{96}\left(\dac{\alpha_1}{\alpha_3}-\dac{9(1-5w)}{2\alpha_2}\varkappa\varepsilon_0\right)\sigma_2+\dac{1-3w}{384\alpha_3}\varkappa\varepsilon_0,\end{array}$$
where $$\sigma_s\equiv\sum_i H_i^s.$$ The possibility of a solution in an arbitrary order is thus
far not evident. It depends on the dimensionality, which order one should be restricted to,
but we so far do not see any reasoning that would enable us to judge on the dimensionality.

Thus the main results of the work are as follows: We have proved that a non-isotropizing expansion
of the Universe is possible in different orders of Lovelock's theory. For an arbitrary order,
we have obtained a generalization of the Jacobs solution, and for the third order, a more
attractive exponential solution. It makes possible to reconcile the Lovelock theory with
the non-observability of extra dimensions.

\end{document}